\documentclass[aps,prb,twocolumn,groupedaddress,showpacs]{revtex4-1}
\usepackage{graphicx}
\usepackage{xcolor}

\begin{document}

\title{Resonance-state-induced superconductivity at high Indium contents in In-doped SnTe}

\author{Neel Haldolaarachchige, Quinn Gibson, Weiwei Xie, Morten Bormann Nielsen, Satya Kushwaha and R.~J.~Cava}
\affiliation{Department of Chemistry, Princeton University, Princeton, New Jersey 08544, USA}

\begin{abstract}  
We report a reinvestigation of superconducting Sn$_{1-x}$In$_{x}$Te at both low and high In doping levels. Analysis of the superconductivity reveals a fundamental change as a function of \textit{x}: the system evolves from a weakly coupled to a strongly coupled superconductor with increasing indium content. Hall Effect measurements further show that the carrier density does not vary linearly with Indium content; indeed at high Indium content, the samples are overall \textit{n}-type, which is contrary to expectations of the standard picture of In$^{1+}$ replacing Sn$^{2+}$ in this material. Density functional theory calculations probing the electronic state of In in SnTe show that it does not act as a trivial hole dopant, but instead forms a distinct, partly filled In 5\textit{s} - Te 5\textit{p} hybridized state centered around E$_F$, very different from what is seen for other nominal hole dopants such as Na, Ag, and vacant Sn sites. We conclude that superconducting In-doped SnTe therefore cannot be considered as a simple hole doped semiconductor.

\end{abstract}

\pacs{--------}

\maketitle
\newcommand{\angstrom}{\mbox{\normalfont\AA}}

\section{Introduction}
SnTe and other II-VI materials with the rock-salt structure type have long attracted attention as models of small band gap semiconductors, and have been of renewed recent interest due to the discovery of topological crystalline insulators.~\cite{snteti1, snteti2} Hole-doped SnTe is also often considered a model system for superconductivity in a rock-salt type small band gap semiconductor.~\cite{tcvsn_hightc, tcvsn2, tcvsn3, tcvsn4, tcvsn5, sn1-xte_sc} It is known, however, that SnTe shows very different superconducting T$_{c}$'s when self-hole-doped (\textit{i.e.} Sn$_{1-\delta}$Te), and chemically doped (\textit{i.e.} Sn$_{1-x}$In$_{x}$Te), to the same hole-densities.~\cite{sn1-xte_sc, snte_fisher, snte_ando, snte_gu} Moreover, Indium-doped SnTe maintains the cubic rock-salt structure type up to very high Indium contents (about 50\%). At such high indium levels, the Sn$_{1-x}$In$_{x}$Te system can no longer be considered as a doped semiconductor because, at such high \textit{x}, normal charge balance rules based on Sn$^{2+}$, In$^{1+}$ and Te$^{2-}$ are strongly violated; \textit{i.e.} for Sn$_{0.5}$In$_{0.5}$Te the In content is an order of magnitude too high for that to be a reasonable model for the system. These observations raise the fundamental question: "What is the real nature of Indium in Sn$_{1-x}$In$_{x}$Te?"

Many reports on this materials system postulate that In$^{1+}$ replaces Sn$^{2+}$. This behavior has been well supported experimentally up to about a 9\% Indium doping level and a good correlation is found between the chemical In content and the experimentally observed hole carrier densities.~\cite{snte_ando} (assuming that each In$^{1+}$ donates one hole to the system when substituting for Sn$^{2+}$). It has also been reported that the superconducting T$_{c}$ of Sn$_{1-x}$In$_{x}$Te continues to increase up to a very high level of doping (50\% of In). Unlike the case for the lower doping levels, however, there do not appear to be any reports of a correlation between In content and carrier density in this composition regime; only T$_{c}$'s and upper critical fields are presented. This therefore raises another question: "If Sn$_{1-x}$In$_{x}$Te is the model system for hole-doping induced superconductivity in a rock salt structure semiconductor, then does the superconducting T$_{c}$ scale with hole-density, even at high doping levels?'' 

To test these two questions, we have revisited Indium-doped SnTe. Our experimental investigations have revealed unanticipated new details of the electronic behavior of the system, leading us to suggest that Indium is not a trivial dopant in SnTe. Motivated by the experiments, analysis of our electronic band structure calculations indicates that unlike the case for other monovalent dopants, the In(\textit{s}) states in Sn$_{1-x}$In$_{x}$Te are prevalent at the Fermi level, supporting a resonant-level-type model in which Indium has a mixed oxidation state in the system, \textit{i.e.} that it has partially filled 5\textit{s} states and thus is neither In$^{1+}$ nor In$^{3+}$. Indium as a resonant level dopant has previously been proposed in SnTe~\cite{snte2013pnas, sntepbte, hermans2012}  (along with the analogous Tl-doped PbTe system~\cite{pbtlte}). Here we strongly support this viewpoint - we show from both experimental studies and DFT calculations that superconducting Sn$_{1-x}$In$_{x}$Te cannot be viewed as a simple hole doped semiconducting material.

\section{Experiment and Calculation}
Polycrystalline samples of Sn$_{1-x}$In$_{x}$Te were prepared by a single step solid state reaction method, starting with ultra-high purity (5N, 99.999\%) elemental Sn, In and Te. The starting materials were placed in quartz glass tubes and sealed under vacuum. The tubes were heated (180~$^{0}$C per hour) to 1100 $^{0}$C and held at that temperature for about 5 hrs. They were then rapidly cooled to 850~$^{0}$C, and held there about 10 hrs after which they were again rapidly cooled to room temperature. The purity and cell parameters of the samples were evaluated by powder X-ray diffraction (PXRD) at room temperature on a Bruker D8 FOCUS diffractometer (Cu$~K_{\alpha}$) and unit cell parameters were determined by least squares fitting of the peak positions with the MAUD program.~\cite{maud} Further investigation of the sample purity was done with Energy Dispersive Spectroscopy (EDS) by using a FEI XL30 FEG-SEM system equipped with an EVEX EDS. EDS studies on the In-doped samples indicated that the nominal and actual In-doping concentrations are closely matched. Therefore, the nominal concentrations are used throughout this manuscript. Single crystal X-ray diffraction measurements on a single crystal of Sn$_{0.6}$In$_{0.4}$Te extracted from the characterized polycrystalline sample were carried out at 100 K on a Bruker Apex II diffractometer with Mo radiation. Details of the data collection and analysis are found in the supplementary information file.

The electrical resistivities were measured using a standard four-probe method with an excitation current of 10 mA; small diameter Pt wires were attached to the samples using a conductive epoxy (Epotek H20E). Data were collected from 300 - 0.4 K in magnetic fields up to 5 T using a Quantum Design Physical Property Measurement System (PPMS) equipped with a $^{3}$He cryostat. Hall Effect measurements were similarly made with a 4-wire configuration geometry, in an applied field of $\pm$~1~T to subtract off the possible longitudinal resistive contribution. Specific heats were measured between 0.4 and 50 K in the PPMS, using a time-relaxation method, at 0 and 5 Tesla applied magnetic fields. The magnetic susceptibilities were measured in a DC field of 10 Oe; the samples were cooled down to 1.8 K in zero-field, the magnetic field was then applied, and the sample magnetization was followed on heating to 6 K [zero-field-cooled (ZFC)], and then on cooling to 1.8 K [field-cooled (FC)] in the PPMS.

The electronic structure calculations were performed by density functional theory (DFT) using the WIEN2K code with a full-potential linearized augmented plane-wave and local orbitals [FP-LAPW + lo] basis~\cite{blaha2001,sign1996,madsen2001,sjosted2000} together with the PBE parameterization~\cite{perdew1996} of the GGA, with spin orbit coupling (SOC). The plane-wave cutoff parameter R$_{MT}$K$_{MAX}$ was set to 7 and the Brillouin zone was sampled by 100 k points. Supercells were created to accommodate the dopant impurity atoms. The 3\% doping level was simulated with a primitive cubic unit cell, and the 12\% doping level was simulated with a face centered cubic unit cell, both with one impurity atom per unit cell, placed as a substitution on the Sn site.

\section{Materials Characterization}

\begin{figure}[t]
  \centerline{\includegraphics[width=0.5\textwidth]{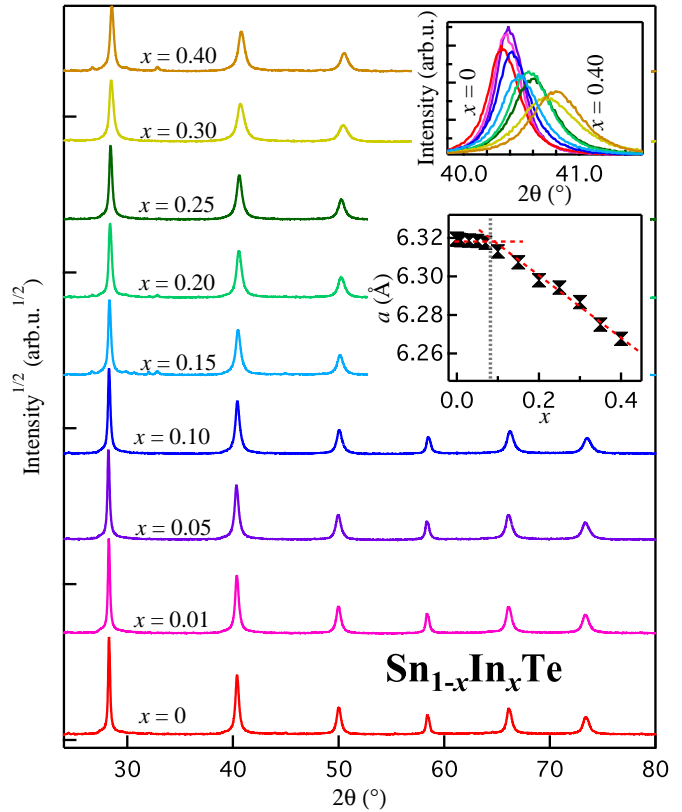}}
  \caption
    {
      (Color online) Powder X-ray diffraction (PXRD) data for the Indium doped SnTe. Upper left inset shows systematic peak shift with increasing In doping and lower left inset shows lattice parameter variation of Sn$_{1-x}$In$_{x}$Te samples in the range of $0 \le x \le 0.4$.
    }
  \label{Fig1}
\end{figure}

Fig.~\ref{Fig1} shows the powder X-ray diffraction (PXRD) patterns obtained for the Sn$_{1-x}$In$_{x}$Te materials. The patterns are a good match to the NaCl-type structure of SnTe (\emph{x} = 0, \emph{Fm-3m}, \emph{a} = 6.32 \AA ), with a composition-dependent unit cell parameter shift. The main panel of Fig.~\ref{Fig1} shows that a single phase material with the cubic NaCl structure is maintained in the Indium doped samples up to a very high doping level (40\%). The upper right inset of Fig.~\ref{Fig1} shows an expanded view of the diffraction peak near 2$\theta$ = 40 degrees. Systematic peak shifts are observed, indicating a systematic shrinking of the unit cell as a function of In content. This behavior is further highlighted by plotting the lattice parameter variation with composition (see lower right inset of the Fig.~\ref{Fig1}), which shows continuous shrinking of the unit cell. A similar shrinkage of the lattice parameter with indium content was found in previous studies of Sn$_{1-x}$In$_{x}$Te up to the 50\% doping level.~\cite{snte_gu} The data support the widely held belief that In systematically replaces the Sn atoms in SnTe to create a single phase NaCl-type structure material. The composition dependence of the lattice parameter variation (see lower right inset of the Fig.~\ref{Fig1}), however, shows that $\left( \frac{da}{dx}\right)$ changes slope at around the 9\% doping level, indicating that there are changes in the system near 9\% In content. This structural change around 9-10 \% indium doping is well correlated with other observations of the electronic properties, as described later in this paper. 

Due to the extremely high In content that can be substituted for Sn in SnTe, and the changes in the cell parameter vs. composition behavior described above, we considered the possibility that the crystal structure of Sn$_{1-x}$In$_{x}$Te for high \textit{x} might not be a simple NaCl type. To maintain the charge neutrality expected for semiconductors, for example, a highly defective material of composition Sn$_{1-x}$In$_{x}$Te$_{1-(0.5x)}$ could conceivably be formed at high \textit{x}. Alternatively, even if the material is essentially stoichiometric at Sn$_{1-x}$In$_{x}$Te, at very high \textit{x} values some or all of the In could be found in tetrahedral interstitial positions in the rocksalt framework, rather than substituting on the octahedral site occupied by Sn, in other words the structural formula could be [Sn(octahedral)$_{1-x}$In(tetrahedral)$_{x}$]Te for high \textit{x}. To test these possibilities, a very careful single crystal structure determination was performed on a single crystal of formula Sn$_{0.6}$In$_{0.4}$Te at 100 K. The crystal was found, to high precision, to have the perfect, stoichiometric rocksalt structure with all ideal atomic sites fully occupied and In simply substituting for Sn. Further, there were no displacements of the atomic positions from the high symmetry ideal rocksalt structure positions. Thus Sn$_{1-x}$In$_{x}$Te is structurally and chemically exactly as it has been assumed to be in previous studies: its crystal structure can confidently be assigned to a simple, stoichiometric NaCl-type. The details of the data collection, structural analysis procedure, and results can be found in the supplementary information.      

\begin{figure}[t]
  \centerline{\includegraphics[width=0.5\textwidth]{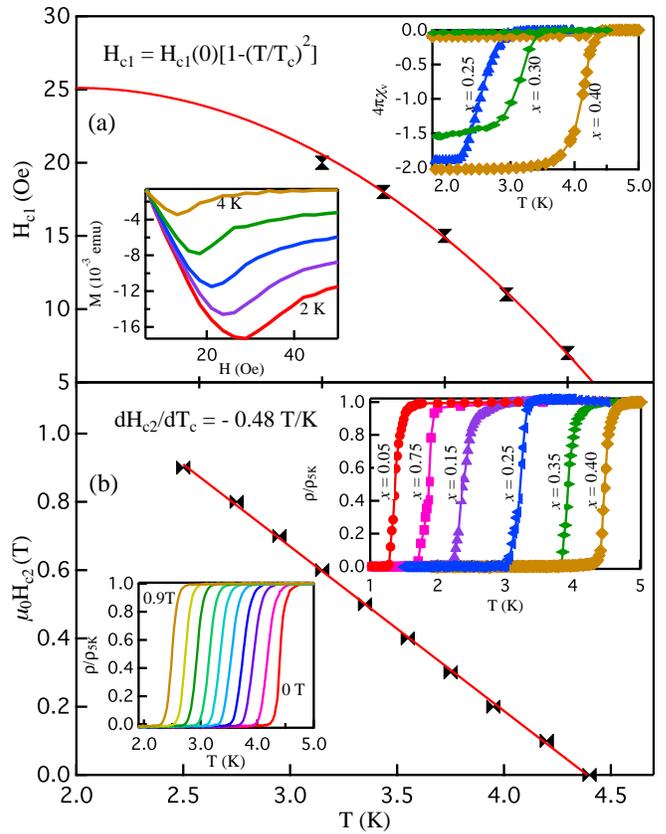}}
  \caption
    {
      (Color online)  (a) Lower critical field and (b) upper critical field analysis of 40\% Indium doped SnTe sample. The upper left inset of (a) shows the magnetic susceptibility data and the lower right inset of (a) shows the magnetization as a function of magnetic field at temperatures below the zero field superconducting temperature. The main panel of (a) shows the conventional fitting for determining the lower critical field. The upper inset of (b) shows the resistivity as function doping and shows the resistivity as a function of temperature at different applied magnetic fields. (d) shows the WHH fitting to the upper critical field values.(a) Resistivity, (b) magnetic susceptibility and (c) carrier density behavior as function of temperature for Sn$_{1-x}$In$_{x}$Te.
    }
  \label{Fig2}
\end{figure}

\begin{figure}[t]
  \centerline{\includegraphics[width=0.5\textwidth]{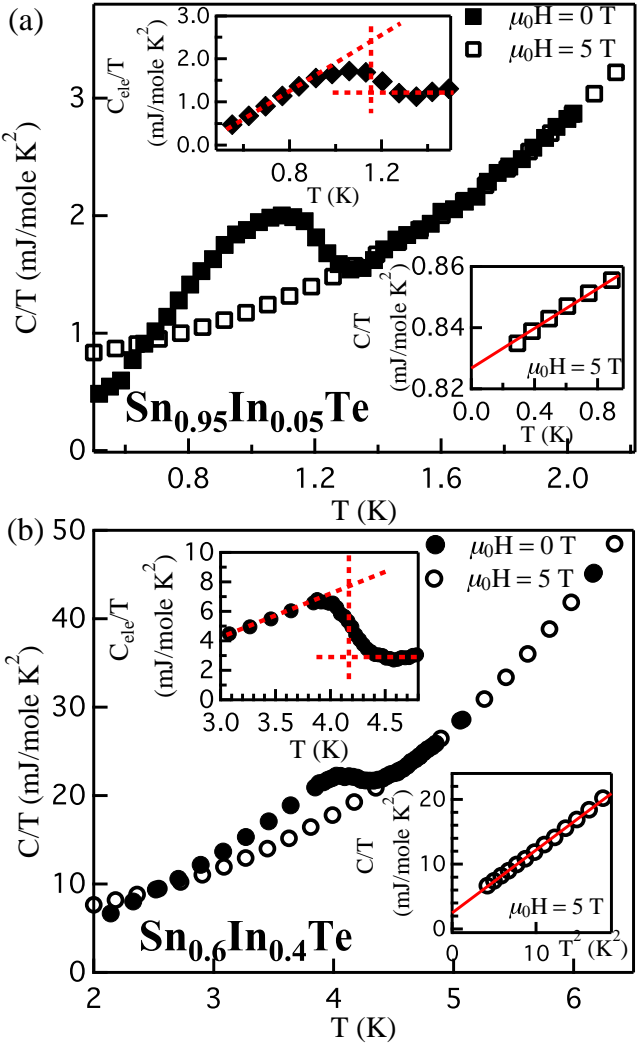}}
  \caption
    {
      (Color online) Heat capacity analysis of (a) 5\% and (b) 40\% Indium doped SnTe samples. The upper left insets of both panels show the superconducting phase transitions in the electronic heat capacity, and the lower right insets show the heat capacity at 5 Tesla applied magnetic field.
    }
  \label{Fig3}
\end{figure}

The upper right inset of Fig.~\ref{Fig2}.(a) shows the magnetic susceptibility characterization of the superconducting transitions of Sn$_{1-x}$In$_{x}$Te. The superconducting shielding can be observed in the zero-field-cooled (ZFC-shielding) and field-cooled (FC-Meissner) data in the figure. The large volume fractions observed confirm the bulk superconductivity, and a systematic increase of superconducting T$_{c}$ is easily observed in the high In content range of 20-40\%. Very similar superconducting transition temperature values can be observed in both resistivity and magnetization data on the 20-40\% Indium doped samples. This is a good indication of the homogeneous quality of the polycrystalline samples studied.

The main panel of Fig.~\ref{Fig2}.(a) shows the detailed analysis of the lower critical field behavior for Sn$_{0.6}$In$_{0.4}$Te. The 40\% Indium doped material is selected for this study because it has the highest T$_{c}$ in the series of samples prepared here. The lower left inset of Fig.~\ref{Fig2}.(a) shows the magnetization as a function of applied magnetic field at several temperatures below the zero field superconducting T$_{c}$. The behavior confirms the type-II superconductivity. The solid line in the figure shows the fitting to the conventional formula $H_{c1}(T)=H_{c1}(0)\left[1-\left(\frac{T}{T_{c}}\right)^{2}\right]$. The lower critical field can be extracted as $H_{c1}(0)$ = 21 Oe.~\cite{tinkham}

The upper right inset of Fig.~\ref{Fig2}.(b) shows the resistivity variation for Sn$_{1-x}$In$_{x}$Te. Pure SnTe shows metallic-like behavior $\left( \frac{d\rho}{dT} > 0\right) $ with \textit{p}-type carrier density (10$^{20}$ cm$^{-3}$) (not shown here), which agrees well with previously published data.~\cite{snte_fisher, snte_ando} The material becomes superconducting immediately with small amounts of Indium doping. Our data agrees well with the data reported in the literature. The upper right inset of Fig.~\ref{Fig2}(b) shows that the superconducting T$_{c}$ is in the 1-2 K range at low doping levels of Indium ($x < 0.1$) and increases linearly with In content. Also, the superconducting T$_{c}$ increases linearly as a function of doping at higher doping levels ($x > 0.1$). There is a very clear change in the slope of T$_{c}$ vs. \textit{x} that can be observed at around the 9-10\% Indium doping level. This behavior is consistent with the observed anomaly in lattice parameter variation shown in Fig.~\ref{Fig1}.

Continuing the characterization of the superconductivity for the highly doped material, the main panel of Fig.~\ref{Fig2}.(b) shows analysis of upper critical field of Sn$_{0.6}$In$_{0.4}$Te. The width of the superconducting transition decreases systematically with increasing magnetic field (see lower left inset of Fig.~\ref{Fig2}.(b)). Selecting the 50$\%$ normal state resistivity point as the transition temperature, we estimate the orbital upper critical field, $\mu_{0}H_{c2}$(0), from the Werthamer-Helfand-Hohenberg (WHH) expression 
$\mu _{0}H_{c2}(0)=-0.693~T_{c}\frac{dH_{c2}}{dT}\vert_{T=T_{c}}$. This expression is widely used to calculate the upper critical field of a variety of superconductors including intermetallic, heavy metal and oxide based materials.~\cite{amar2011, neelkwo3, neelbabi3, neelcair2} A very linear relationship is observed in the main panel of Fig.~\ref{Fig2}(b) between $\mu _{0}H_{c2}$ and $T_{c}$. The slope $\left( \frac{d \mu _{0}H_{c2}}{dT_{c}} = - 0.48~T/K\right) $ is used to calculate $\mu _{0}H_{c2}(0)=$~1.46~T. This value of $\mu _{0}H_{c2}(0)$ is smaller than the weak coupling Pauli paramagnetic limit $\mu _{0}H^{Pauli} = 1.82~T_{c} = 7.56$ T for this system. 

The upper critical field value $\mu _{0}H_{c2}(0)$ can be used to estimate the coherence length $\xi (0)=\sqrt{\Phi _{0}/2\pi H_{c2}(0)} = 150$~\AA, where $\Phi _{0}=\frac{hc}{2e}$ is the magnetic flux quantum.~\cite{clogston1962, werthamer1966} 
Similarly, from the relation of $H_{c1}(0)=\frac{\phi_{0}}{4\pi \lambda^{2}}ln\frac{\lambda}{\xi}$, we find the magnetic
penetration depth $\lambda(0) = 5000$ \AA.
A Ginzburg-Landau parameter $\kappa = \frac{\lambda}{\xi} = 33$ is then calculated. 
Using these parameters and the relation of $H_{c2}(0)H_{c1}(0) = H_{c}(0)^{2}[ln \kappa(0) + 0.08]$, the
thermodynamic critical field H$_{c}$(0) was found to be 0.85 mT. All the superconducting parameters determined for Sn$_{0.6}$In$_{0.4}$Te are summarized in Table.~\ref{tab:1}.

Fig.~\ref{Fig3} shows the analysis of the superconducting transition by specific heat measurements for \textit{p}-type Sn$_{0.95}$In$_{0.05}$Te and \textit{n}-type Sn$_{0.6}$In$_{0.4}$Te. 
The main panel of Fig.~\ref{Fig3}.(a) shows $\frac {C}{T}$ as a function of $T$ for Sn$_{0.95}$In$_{0.05}$Te, characterizing the specific heat jump at the thermodynamic transition. This jump is completely suppressed under a 5 T applied magnetic field. The superconducting transition temperature $T_{c}$ = 1.18 K is shown in the upper left inset of Fig.~\ref{Fig3}.(a), as extracted by the standard equal area construction method. 
We find that the low temperature normal state specific heat can be well fitted with $\frac{C}{T} = \gamma _{n} + \beta T^{2}$,
where $\gamma _{n} T$ represents the electronic contribution in the normal state and $\beta T^{3}$ describe the lattice-phonon contributions to the specific heat. 
The solid line in the lower right inset in Fig.~\ref{Fig3}.(a) shows the fitting; the electronic specific heat coefficient $\gamma _{n} = 0.82 \frac{mJ}{mol~K^{2}}$ and the phonon contribution $\beta = 0.45 \frac{mJ}{mol~K^{3}}$ are extracted from the fit. 
The value of $\gamma _{n} $ for this \textit{p}-type superconductor is consistent with the previously reported values.~\cite{snte_ando, snte_fisher} 

The main panel of Fig.~\ref{Fig3}.(b) shows $\frac {C}{T}$ as a function of $T$ for Sn$_{0.6}$In$_{0.4}$Te, characterizing the specific heat jump at the thermodynamic transition. This jump is completely suppressed under a 5 T applied magnetic field. The superconducting transition temperature $T_{c}$ = 4.2 K is shown in the upper left inset of Fig.~\ref{Fig3}.(b), as extracted by the standard equal area construction method. 
We find that the low temperature normal state specific heat can be well fitted with $\frac{C}{T} = \gamma _{n} + \beta T^{2}$,
where $\gamma _{n} T$ represents the electronic contribution in the normal state and $\beta T^{3}$ describes the phonon contribution to the specific heat. 
The solid line in the lower right inset in Fig.~\ref{Fig3}.(b) shows the fitting; the electronic specific heat coefficient $\gamma _{n} = 2.47 \frac{mJ}{mol~K^{2}}$ and the phonon contribution $\beta = 0.97 \frac{mJ}{mol~K^{3}}$ are extracted from the fit. 
The value of $\gamma _{n} $ for this highly In-doped \textit{n}-type superconductor is much higher than that of the low level Indium doped \textit{p}-type sample.
The specific heat data are a clear indication that some very specific difference is present between the low and high level Indium doped samples.

The ratio $\frac{\Delta C}{\gamma T_{c}}$ can be used to measure the strength of the electron-phonon coupling.~\cite{padamsee1973} The specific heat jump $\frac{\Delta C}{T_{c}}$ for Sn$_{0.95}$In$_{0.05}$Te is about 1.2 $\frac{mJ}{mol~K^{2}}$, which results in the value of $\frac{\Delta C}{\gamma~T_{c}}$ of 1.45. This value is about the same as the BCS prediction for weakly electron-phonon coupled superconductors and also agrees with previously reported values of low level of Indium doped samples.~\cite{snte_fisher, snte_ando}
However, The specific heat jump $\frac{\Delta C}{T_{c}}$ for the sample of Sn$_{0.6}$In$_{0.4}$Te is about 4.9 $\frac{mJ}{mol~K^{2}}$, which results in a value of $\frac{\Delta C}{\gamma~T_{c}}$ of 1.98. 
This is higher than that of the weak-coupling limit for conventional BCS superconductors. Therefore, the results suggest that Sn$_{0.6}$In$_{0.4}$Te is a strongly electron$-$phonon coupled superconducting system. The observed values of $\frac{\Delta C}{\gamma~T_{c}}$ show that the low and high doping levels of Indium in SnTe make two distinct types of superconductors. 

In a simple Debye model for the phonon contribution to the specific heat, the $\beta_{1}$ coefficient is related to the Debye temperature $\Theta _{D}$ through $\beta = nN_{A}\frac{12}{5}\pi ^{4}R\Theta _{D}^{-3}$, where $R = 8.314~\frac{J}{mol~K}$, $\textit{n}$ is the number of atoms per formula unit and $N_{A}$ is Avogadro\textquoteright s number. The calculated Debye temperatures are thus 204 K and 162 K for 5\% and 40\% Indium doped samples. These values of the Debye temperatures are similar to the previously reported values on chemically doped SnTe, PbTe and related systems.~\cite{snte_fisher, pbtlte, pbtlinte}
An estimation of the strength of the electron-phonon coupling can be derived from the McMillan formula
$\lambda _{ep} = \frac{1.04 + \mu ^{*} ln\frac{\Theta _{D}}{1.45T_{c}}}{(1-0.62\mu ^{*}) ln\frac{\Theta _{D}}{1.45T_{c}}-1.04}$.
The McMillan model contains the dimensionless electron-phonon coupling constant $\lambda _{ep}$, which, in the Eliashberg theory, is related to the phonon spectrum and the density of states.~\cite{mcmillan1968, poole1999}  This parameter $\lambda _{ep}$ represents the attractive interaction, while the second parameter $\mu ^{*}$ accounts for the screened Coulomb repulsion.
Using the Debye temperature $\Theta _{D}$ and the critical temperature $T_{c}$, and making the common assumption that $\mu ^{*} = 0.15$,~\cite{mcmillan1968} the electron-phonon coupling constants ($\lambda _{ep}$) are 0.52 and 0.79 for 5\% and 40\% Indium doped samples. Thus, our characterization of the superconducting transitions supports the conclusion that the 5\% Indium doped sample can be categorized as a weakly coupled superconductor and the 40\% Indium doped sample can be categorized as a strongly coupled superconductor. 

The value of $\gamma$ extracted from the measured specific heat data corresponds to a normalized electronic density of states at the Fermi energy N($E_{F}$). The following values for the density of states, 0.44 and 1.17 states/(eV f.u.) (f.u. stands for formula unit) for 5\% and 40\% Indium doped samples are thus estimated from the relation 
$ \gamma = \pi ^{3/2} k_{B}^{2} N(E_{F}) (1+ \lambda _{ep})$. The value of N($E_{F}$) for the 5\% In-doped samples is consistent with previous reports for low In doping levels but N($E_{F}$) for the 40\% doped sample is much higher, consistent with the fact that that higher In content samples show much higher superconducting T$_{c}$'s than the lower In content samples.

\begin{figure}[t]
  \centerline{\includegraphics[width=0.5\textwidth]{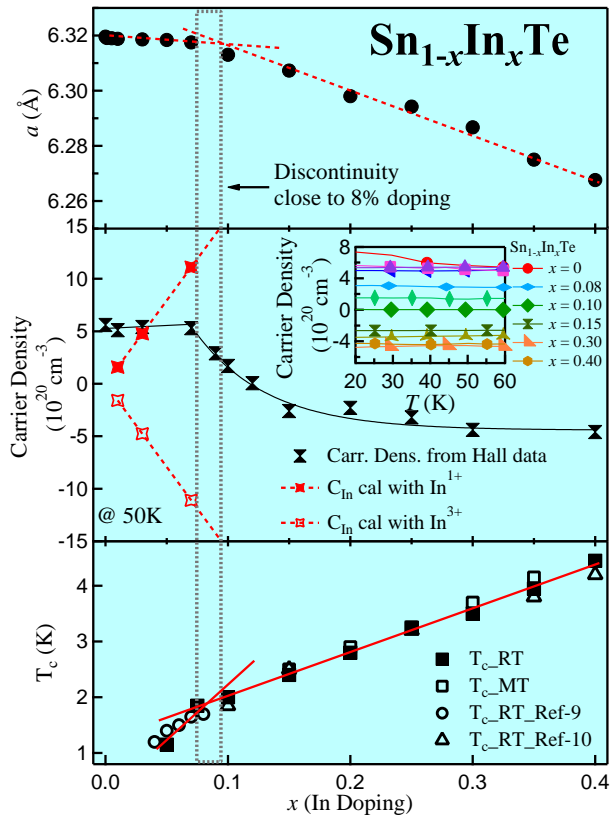}}
  \caption
    {
      (Color online) the (a) Lattice parameter variation, (b) carrier density variation and volume density of Indium atoms in the doped system, and (c) superconducting transition temperature variation, as a function of fractional indium content in the cubic In-doped SnTe crystal system. The inset of (b) shows the raw data for the carrier density determination at low T. Solid lines in all panels are guides to the eye.
    }
  \label{Fig4}
\end{figure}

Fig.~\ref{Fig4} (a, b and c) is a summary that shows the lattice parameter variation, carrier density variation, and superconducting temperature variation as a function of Indium doping in SnTe. It can clearly be seen that there is a change at the 9-10\% Indium doping level in Fig.~\ref{Fig4}.(a) that is well correlated with the \textit{p} to \textit{n} type crossover that is seen in Fig.~\ref{Fig4}(b). At low doping levels of Indium ($\textit{x} < 10\%$) the system remains \textit{p}-type, which agrees well with previous reports. However, when the doping level goes beyond a critical doping level (\textit{x} = 10\%), the system shows anomalous behavior of the carrier density. At higher doping levels of Indium ($\textit{x} > 10\%$), the system changes to \textit{n}-type and the composition dependence of the carrier density saturates quickly. This change should be connected to some kind of Fermi surface reconstruction; 
the behavior is not consistent with the conventional picture of hole-doping through In$^{1+}$ substitution for Sn$^{2+}$, which we find to be true only up to about 9\% Indium doping. Fig.~\ref{Fig4}(b) also shows the volume density expected by assuming that every dopant Indium atom donates a hole, or an electron, into the SnTe unit cell. In such cases the carrier density as a function of dopant concentration should linearly increase in a positive direction for In$^{1+}$ substitution or increase in a negative direction for In$^{3+}$ substitution. However, the behavior in this system is much more complicated than that. In doping results in the unexpected suppression of the \textit{p}-type carrier density at high \textit{x} and, that at a very high doping level, the carrier type changes to \textit{n}-type. More detailed experimental studies of single crystals in this composition regime to characterize this crossover in more detail would be an interesting avenue for future work.

\begin{table}[t]
\caption{Superconducting Parameters of the cubic \textit{p}-type and \textit{n}-type Indium doped SnTe systems}. 
  \centering  
  \begin{tabular}{ lc   c   c   }
  \hline \hline 
    Parameter & Units &  Sn$_{0.95}$In$_{0.05}$Te & Sn$_{0.6}$In$_{0.4}$Te     \\ \hline  \hline  
    $T_{c}$ & K & 1.18 & 4.2 \\
    $\frac{dH_{c2}}{dT}\vert _{T=T_{c}}$ & $T~K^{-1}$ &   & -0.48 \\
    $\mu _{0}H_{c1}(0)$ & Oe &   &  21  \\
    $\mu _{0}H_{c2}(0)$ & T &    &  1.46 \\
    $\mu _{0}H^{Pauli}$ & T & 2.23  &  7.81 \\
    $\mu _{0}H (0)$ & mT &   &  0.85 \\
    $\xi (0)$ & \AA &   &  150.2 \\
    $\lambda (0)$ & \AA &   &  5000 \\
    $\kappa (0)$ & \AA &  &  33.28 \\
    $\gamma (0)$ & $\frac{mJ}{mol~K^{2}}$ & 0.94  & 2.47  \\
    $\frac{\Delta C}{\gamma T_{c}}$ & & 1.27   &  1.98 \\
    $\Theta _{D}$ & K & 204   &  162  \\
    $\lambda _{ep}$ &  & 0.52  &  0.79  \\
    $N(E_{F})$ & $\frac{eV}{f.u.}$ &  0.44  &  1.17   \\ \hline \hline
  \end{tabular}
  \label{tab:1}
\end{table}

It can be observed that the superconductivity emerges immediately with Indium doping into SnTe Fig.~\ref{Fig4}(c). The superconducting T$_{c}$ continuously increases as a function of doping. However, there is a clear deviation at 9-10\% Indium doping level in the rate of increase of the superconducting temperature as a function of doping; $\left( \frac{dT_{c}}{dx_{In}}\right) $ is higher at lower doping levels of Indium ($x < 0.1$) and becomes smaller when $x > 0.1$. This point of deviation around 10\% Indium is well correlated with the lattice parameter and carrier density variations. 

\section{Electronic Structure}

\begin{figure*}[h]
  \centerline{\includegraphics[width=1.1\textwidth]{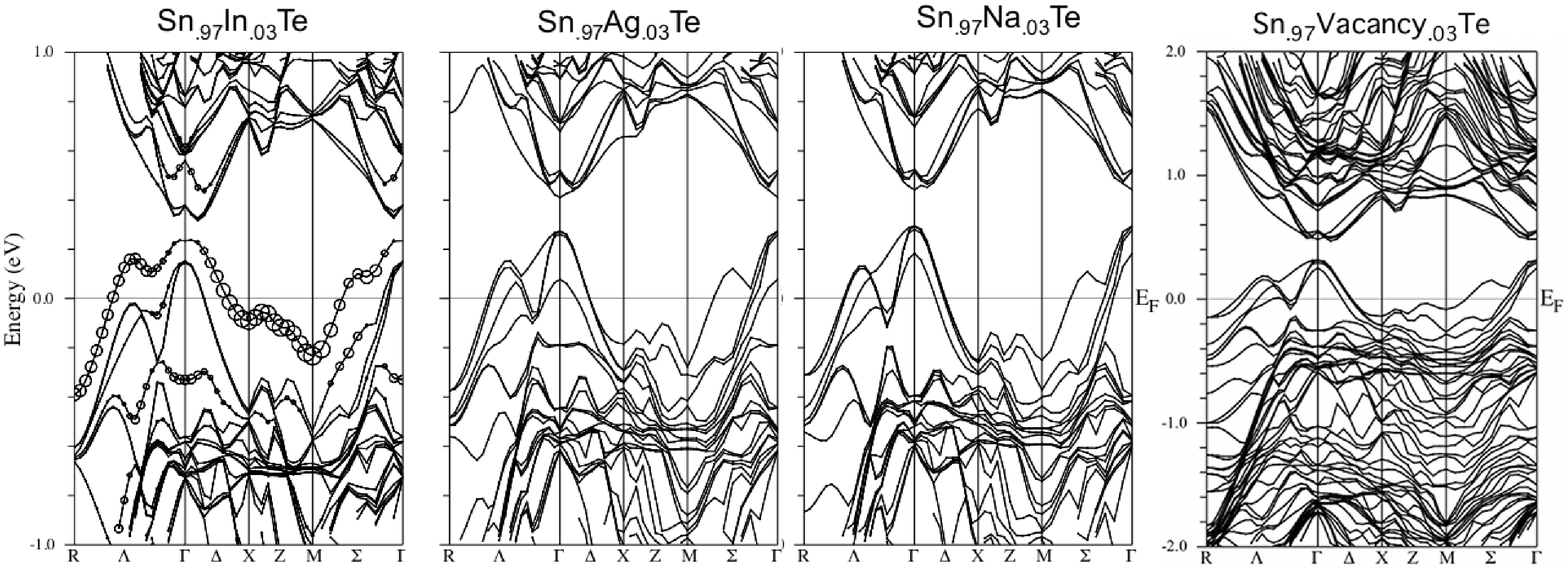}}
  \caption
    {
      (Color online) calculated electronic band structure for 3\% In, Ag, Na and vacancy doped SnTe. The resonant band at E$_{F}$ can be observed only in the In doped case.
    }
  \label{Fig5}
\end{figure*}

Future experimental investigation of the higher In content materials will be of interest, but here we look in more detail at the apparent complexity of the electronic system by performing electronic band structure studies on model materials that simulate the effects of doping in SnTe. In order to theoretically investigate the electronic structure of doped SnTe, DFT calculations were performed on supercells containing different levels of In, Ag, Na and Sn-vacancies; the latter dopants are considered for comparison to the In case. Fig.~\ref{Fig5} shows the electronic band structure for 3\% doped SnTe with In, Ag, Na and vacancy dopants. Quite dramatically, the Ag, Na and Sn-vacancy doped models are qualitatively very similar, with the Fermi energy about 200 meV deep in the valence band. However, the model for the In doped material is qualitatively different. The Fermi energy is less deep in the valence band, and there is a new, distinct in-gap state not seen in the other calculations. The fat bands, which allow the orbital origin of this in-gap state to be determined, show that it is due to the contributions from the In 5\textit{s} orbital. Further analysis shows this band to be composed primarily of In 5\textit{s} and Te 5\textit{p} orbitals. This in-gap bands cuts through the Fermi energy at multiple points, and thus contributes to the electronic properties of the material. This already indicates an unusual doping mechanism. While In is creating holes in the valence band manifold, it is simultaneously creating other electron and hole pockets through the creation of an impurity band centered around E$_F$.

\begin{figure*}[h]
  \centerline{\includegraphics[width=1.1\textwidth]{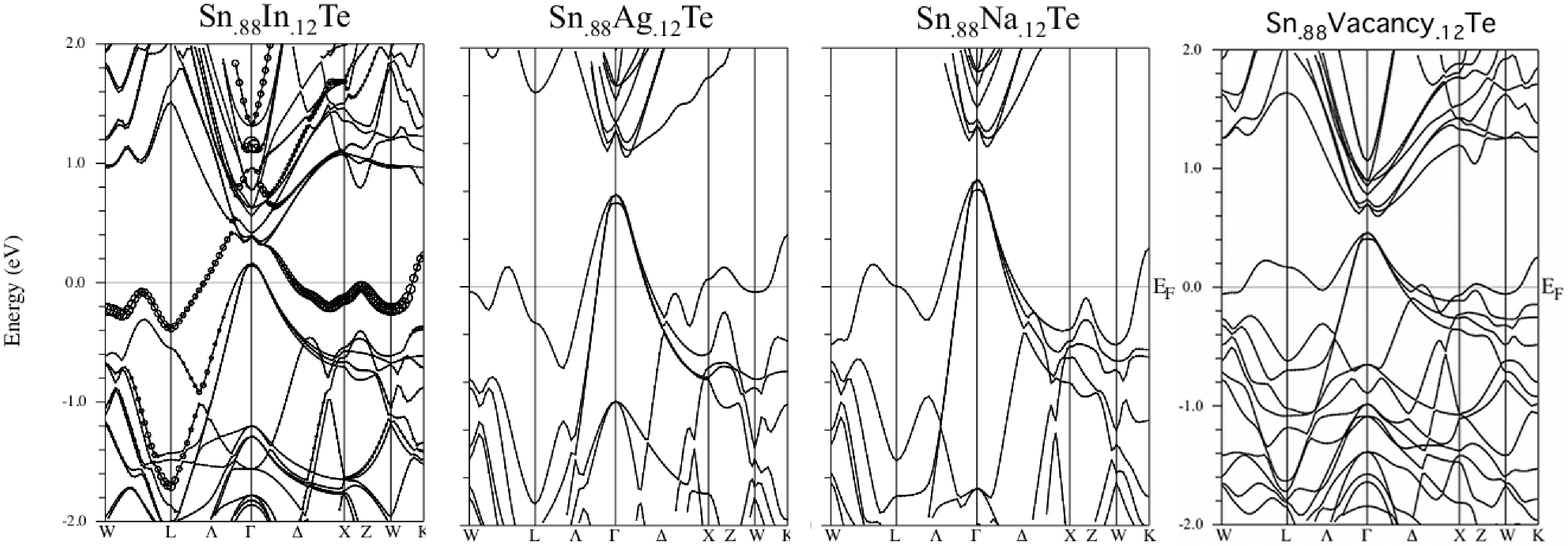}}
  \caption
    {
      (Color online) calculated electronic band structure for 12\% In, Ag, Na and vacancy doped SnTe. The resonant band at E$_{F}$ can be observed only in the In doped case.
    }
  \label{Fig6}
\end{figure*}

This effect is even more pronounced at higher doping levels. Fig.~\ref{Fig6} shows the electronic structures for In, Ag, Na and vacancy doped SnTe at 12 percent doping. Here, again, the Na, Ag and Sn vacancy doped compounds are very similar, whereas the Indium doped compound is qualitatively very different. For one, the Fermi energy is considerably deeper in the valence band for the Ag, Na and defect-doped compounds than for In. Furthermore, the same in-gap state present in the 3 percent doped calculation is present here. Indeed, this band traverses the entire band gap. This compound cannot be considered to be a doped semiconductor, as artificially adding electrons will not bring the Fermi level into a band gap. Finally, the Fermi level is about the same depth in the valence band as is seen for the 3 percent In doping level. While these calculations cannot be directly compared, as the 12 percent and 3 percent calculations needed different unit cell symmetries, this certainly indicates that there is a nonlinear dependence of hole concentration on Indium doping level, at least at high doping levels. This is indeed what is observed experimentally. In fact, as the In impurity band is creating its own Fermi surface, at high doping levels, it may not be meaningful to distinguish between different hole concentrations, as the Fermi surface is now more complex than a single pocket. What is striking is that in both cases, the In dopant is creating an impurity band that is relatively well-separated from the other bulk bands, and which is also not very dispersive in energy. This is characteristic of a resonant level type dopant, as has been discussed in the thermoelectric and related literature.~\cite{snte2013pnas, sntepbte, hermans2012, pbtlte} 

\begin{figure*}[h]
  \centerline{\includegraphics[width=1.1\textwidth]{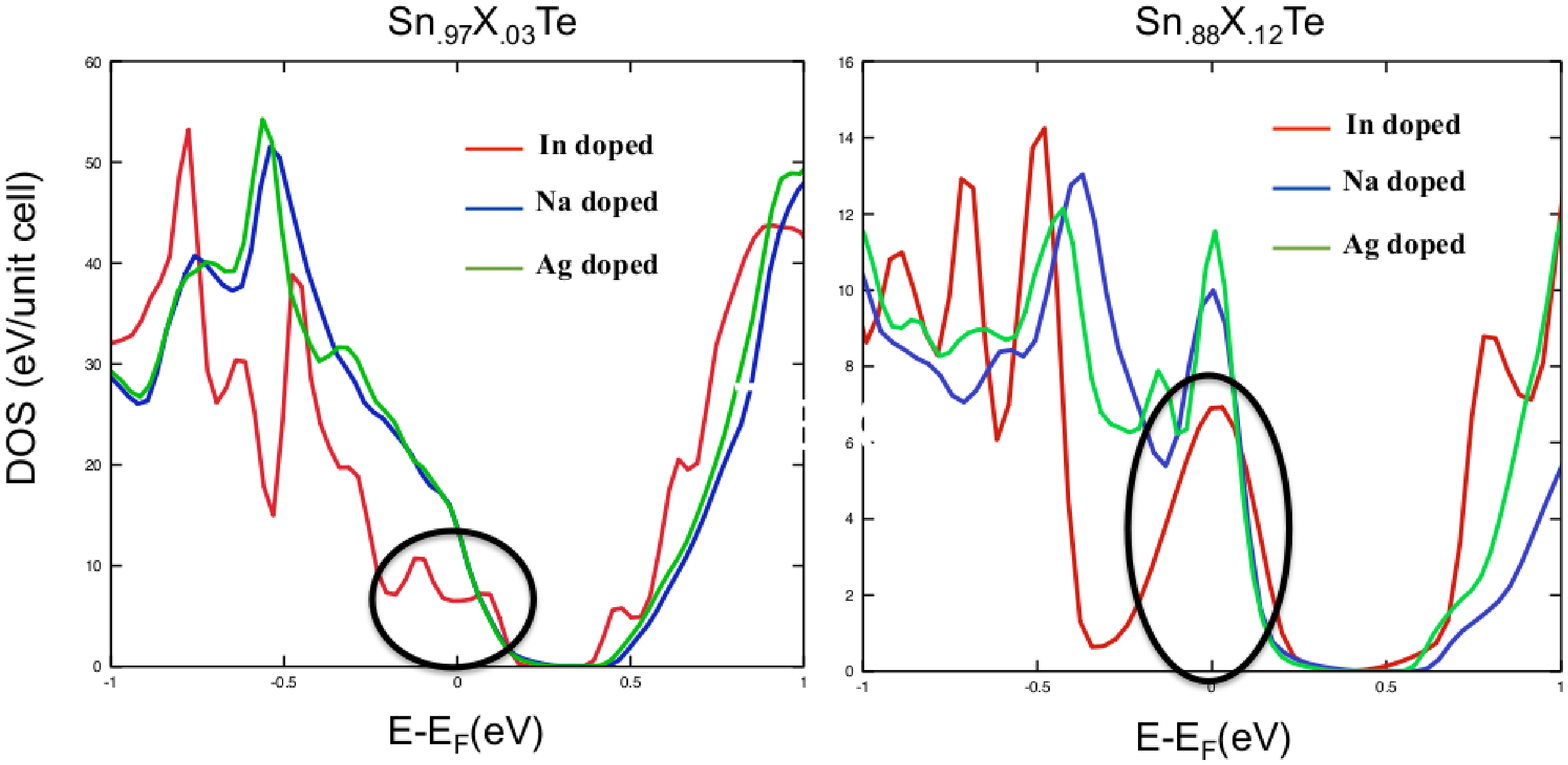}}
  \caption
    {
      (Color online) Density of states as a function of E-E$_{F}$ for (a) 3\% and (b) 12\% In, Ag, and Na doped SnTe systems. Vacancy doping yields a DOS that is very similar to that seen for Ag and Na doping, with vacancies being a two-hole dopant rather than a single-hole dopant. 
    }
  \label{Fig7}
\end{figure*}

The resonant aspect of this impurity state is also apparent in the Density of States. Fig.~\ref{Fig7} shows the DOS for the In, Ag, and Na doped compounds at both the 3 percent and 12 percent doping levels. Again, near E$_F$, the Ag and Na samples resemble each other well, while the In one is different. The In doped compound at 3 percent has a small ''doublet'' peak, which sits right at E$_F$, that is not present in the others. At 12 percent, this has evolved into a tall, well separated peak that is bisected by the Fermi energy. Further analysis show that both of these peaks have nontrivial In 5\textit{s} character, along with Te 5\textit{p} character. This indicates that the impurity state that is so important in In doped SnTe appears to come from a hybridization of In 5\textit{s} and Te 5\textit{p} orbitals. Due to the inert pair effect, it is very unusual to have 5\textit{s} or 6\textit{s} states at the Fermi level (indeed, in SnTe the Sn 5\textit{s} states appear about 5 eV below E$_F$); some well-known compounds that do, such as K doped BaBiO$_3$, exhibit superconductivity.~\cite{kbabio3_sc} It is therefore very likely that the hybrid state created by the In 5\textit{s} and Te 5\textit{p} orbitals play a significant role in the superconductivity. This would explain why In doped SnTe exhibits an order of magnitude higher T$_c$ than self-doped SnTe, even at the same nominal hole concentration. Further calculations indicate that if In were also to occupy an interstitial tetrahedral site rather than replacing an Sn atom, it would act as an \textit{n}-type dopant while still inducing the same in-gap 5\textit{s} state as seen in the previously discussed calculations. This calculation was part of our motivation for the detailed crystallographic study we performed, which showed that In in tetrahedral interstitial sites cannot be found in (Sn,In)Te.

Finally, we note that In has been used as dopant to achieve very high resisitivities in Pb$_{1-x}$Sn$_{x}$Te. This is consistent with the picture of Indium forming an in-gap state. At very low concentrations, this state would be relatively localized, and would pin the Fermi energy to the gap, acting as a carrier concentration buffer, much like Sn does in Sn:Bi$_2$Te$_2$Se.~\cite{pbsnte, satyabitese, andobitese}

\section{Conclusion}
We have shown, through experimental observations and DFT calculations, that Indium doped SnTe cannot be thought of as a simple hole doped semiconductor. The nature of the superconductivity and the carrier type change as a function of Indium doping, going from overall \textit{p}-type to overall \textit{n}-type and from a weakly coupled to a strongly coupled superconductor. Furthermore, the In 5\textit{s} states are shown by theory to be present at the Fermi level and therefore affect the electronic properties. Given that Indium doped SnTe has been studied as a superconducting doped topological crystalline insulator, this work indicates that the nature and influence of the In 5\textit{s} states must be taken into account. This may also suggest that aside from allowing for insulating behavior or improving thermoelectric performance, resonant level dopants can affect superconducting properties. Future studies could elucidate further the nature of resonant level doping in SnTe. It is interesting that a very simple structure type can show both weakly and strongly coupled superconductors. Therefore we also argue that the high In-doped, strongly coupled superconductor in this system is worthy of further investigation.

\section{acknowledgments}
The materials synthesis and physical property characterization of this superconductor were supported by the Department of Energy, division of basic energy sciences, Grant DE-FG02-98ER45706. The single crystal structure determination was supported by the Gordon and Betty Moore Foundation, EPiQS initiative, grant GBMF4412, and the electronic structure calculations were supported by the ARO MURI on topological Insulators, grant W911NF-12-1-0461.

\end{document}